\documentclass{aj}

\long\def\Blind#1\EndBlind{#1}
\begin{document}
\begin{opening}

\title{Why stratification may hurt, \& how much}

\Blind
\author{Chris A.J. \surname{Klaassen}} \&
\author{Andries J. \surname{Lenstra}}

\smallskip\institute{Korteweg-de Vries Institute for Mathematics,
Universiteit van Amsterdam,
Plantage Muidergracht 24, 1018 TV Amsterdam, the Netherlands}

\EndBlind
\runningtitle{Stratification}

\begin{abstract}

There are circumstances under which stratified sampling is worse
than simple random sampling, even if the allocation of the sample
sizes is optimal. This phenomenon
was discovered more than sixty years ago,
but is not
as widely known as one might expect.
We provide it with lower and upper bounds for its badness
as well as with an explanation.
\end{abstract}
\keywords{proportional allocation, replacement,
stratified sampling}
\classification{AMS subject classification (2000)}{%
 Primary: 62D05 (Sampling theory, sample surveys)
Secondary: 62F10 (Parametric inference; point estimation)}
\end{opening}

\def\R{{\rm I\negthinspace R}}\def\P{{\rm I\negthinspace P}}\def\Ltwee{L^2_{{\cal X},\mu}}%

\section{Prologue}
{\em `Stratification is a common technique,'} often necessary, but
also attractive because it {\em `may produce a gain in precision
in the estimates of characteristics of the whole population'}
(Cochran (1963), \S 5.1). In fact, if all sampling is done
with replacement and the sample sizes are proportional to the
strata sizes, stratified random sampling is at least as precise as
simple random sampling. It even approaches perfection as the
homogeneity inside the strata increases, i.e., as the
heterogeneity of the population is more reflected by the
heterogeneity between the strata and less by the heterogeneity
inside the strata. Consequently, the rule of thumb with respect to
stratified sampling is that
it doesn't hurt to try.

In real life,
however,
sampling is {\em without\/} replacement (cf. Cochran
(1963), \S 2.1: {\em `Sampling with replacement is
entirely feasible but except in special circumstances is seldom
used, since there seems little point in having the same unit twice
in the sample'\/})---and without replacement, the rule
of thumb is no longer valid.
Even optimal stratified sampling may hurt then, in
that the corresponding estimator can have a larger variance than
the estimator based on a simple random sample
(cf. Armitage (1947), Cochran (1963) \S5.6,
Evans (1951),
and
Govindarajulu (1999) \S5.5; for the
obscurity of this fact,
please see, for instance,
 the same Govindarajulu (1999) \S5.5
and Wilks (1963), \S10.9).

The intuition might be helped here by realizing that, as
equalities (\ref{truth}) below remind us, {\em not\/} replacing
yields an advantage that is zero for a sample size of~1 and
increases with the sample size. Thus, the advantage is larger for
one sample of size $n>1$ than for~$n$ samples of size~1; cf. the
illustration of Theorem 3 in \S2.

We only consider the simplest possible case, that of a dichotomous
population. For this case, the results in Armitage
(1947), Cochran (1963), and Evans (1951)
are extended to what looks like a quite complete picture.
The simple-is-better effect shows up in more circumstances than previously thought and
is provided with exact bounds for its size.

\section{Lower and Upper Bounds}
Consider an urn containing $N$ balls, of which $pN$ are red and
$(1-p)N$ are black for a~$p\in[0,1]$. We want to estimate~$p$.
One approach is to take a sample of $n$ balls from the urn
and estimate~$p$ by the fraction of the red balls in the sample.
A sample of $n$ balls {\em with replacement\/} is a random member
$(b_1,\ldots,b_n)$ of $({\rm urn})^n$, where all outcomes are equally
likely; a sample of $n$ balls {\em without replacement\/} is the same, except
that the $b_1,\ldots,b_n$ are all different. Let~$X$ and~$Y$ denote
the number of red balls
in a sample of size~$n$ with and without replacement, respectively.
Then  \def\var{{\rm var}}for the fraction estimators $X/n$ and
$Y/n$ for~$p$ the truth of
\begin{equation}\label{truth}
\var {X\over n}={p(1-p)\over n}={N-1\over N-n}\,\var {Y\over n}
\label{well}
\end{equation}
is well known;
it is better never to see the same ball twice.
\def\StratEstWith
 {the Stratification Estimator With Replacement}%
\def\StratEst
 {the Stratification Estimator Without Replacement}%
\def\SimpleEstWith
 {the Simple Estimator With Replacement}%
\def\SimpleEst
 {the Simple Estimator Without Replacement}%
\def\stratestwith
 {the stratification estimator with replacement}%
\def\stratest
 {the stratification estimator without replacement}%
\def\simpleestwith
 {the simple estimator with replacement}%
\def\simpleest
 {the simple estimator without replacement}%

Now suppose the urn consists of $m\geq 2$ disjoint sub-urns, {\em
strata}, each stratum${}_j$ containing~$N_j\geq2$ balls,
$\sum^m_{j=1}N_j=N$, and that for each stratum${}_j$ we know
$N_j/N$ but not its fraction~$p_j$ of red balls. Let $(n_1,
\ldots, n_m)$
be an {\em allocation}, i.e., the~$n_j$ are natural
numbers with $1\leq n_j\leq N_j$ for all~$j$ and $\sum_{j=1}^m n_j
=n$, and let $X_j$, $Y_j$, $j=1\ldots,m$, denote the number of red
balls in a sample of size~$n_j$ from stratum$_j$ with and without
replacement, respectively; then each of
\begin{eqnarray*}
{X\over n}&& (\hbox{\simpleestwith}),\\
\sum_{j=1}^m
{N_j\over N} {X_j\over n_j}&&(\hbox{\stratestwith}),\\
{Y\over n}&&(\hbox{\simpleest}),\\
\hbox{and}\quad\sum_{j=1}^m {N_j\over N} {Y_j\over n_j}&&(\hbox{\stratest})
\end{eqnarray*}
is an
unbiased estimator for~$p$; for their optimality, cf. Neyman
(1934). For both with and without replacement
we want to compare
the variance of the simple estimator to that of
the stratification estimator, i.e.,
$\var (X/n)$ to $\var(\sum_{j=1}^m
(N_j/N) X_j/n_j)$ and $\var(Y/n)$ to $\var(\sum_{j=1}^m (N_j/N)
Y_j/n_j)$,
under the assumption that the~$X_j$ are independent as
well as the~$Y_j$.

It is immediate that if the strata are homogeneous but the whole
population is not, i.e., $p\in(0,1)$ and each $p_j$ is equal
to~0 or~1, then the stratification estimators
are perfect while, with replacement, the simple estimator is not,
and, without replacement, the simple estimator is only
perfect when it is exhaustive, so that 
$0=\var(\sum_{j=1}^m (N_j/ N) X_j/n_j)<\var(X/ n)$ and
$0=\var(\sum_{j=1}^m (N_j/ N) Y_j/n_j)\leq\var(Y/ n)$.

For arbitrary $p_j$, the stratification estimator
$\sum_{j=1}^m (N_j/N) X_j/n_j$ is still not worse than
the simple estimator
$X/n$ as long as the allocation is
{\em proportional}, i.e., $n_j=(N_j/N)n$ for every~$j$,
because in that case
\begin{equation}\label{PropWithRepl}
\var \sum_{j=1}^m {N_j\over N} {X_j\over n_j}=
\var {X\over n} -
   {1\over n}\sum_{j=1}^m {N_j\over N}(p_j-p)^2
\end{equation}
holds, as one easily verifies.
Thus, if all sampling is with replacement and the allocation
is proportional, then
stratified sampling is seen to reduce the variance,
unless all the~$p_j$ are equal.
And where it doesn't help, it doesn't harm either.
However, this reassurance no longer holds as soon as we change  the
allocation:
\begin{thm}\label{WithNotProp}
If $p_1=\cdots =p_m=p\in(0,1)$ and the allocation
is {\em not} proportional, then 
{simple is
better} in that
$$
\var \sum_{j=1}^m {N_j\over N} {X_j\over n_j}
>
\var {X\over n},
$$
i.e., the variance of \stratestwith{}
is greater than the variance of \simpleestwith.
\end{thm}
Nor does it hold if all samples are drawn
{\em without\/} replacement:
\begin{thm}\label{WithoutReplWithoutMalus}
If $p_1=\cdots =p_m=p\in(0,1)$ and $n<N$, then\/
{simple is better} in that
\begin{equation}
\var \sum_{j=1}^m {N_j\over N} {Y_j\over n_j}
>
\var {Y\over n},
\label{Thm2}
\end{equation}
i.e., the variance of \stratest{}
is greater than the variance of \simpleest.
\end{thm}
This is (vi{\em b}) in Armitage (1947) for a dichotomous population,
i.e.,
 for the case where Armitage's `variable~$x$'
has only 2 different values, except that we do not need his
condition that (in the dichotomous case) all the~$N_j$ are equal.
It is also the dichotomous case of what is proved at the end of
\S5.6 in Cochran (1963), except that our condition $p_j=p$ is
replaced by the condition that all the `mean square [errors]
within strata' $p_j(1-p_j)N_j/(N_j-1)$ are equal and larger than
the `mean square [error] among strata' $\sum_{j=1}^m
N_j(p_j-p)^2/(m-1)$. In H\'ajek (1981), the observation after
(20.31) that simple is better if $p_j=p$ only refers to
proportional allocation, not necessarily to all allocations.
%

Under additional conditions inequality (\ref{Thm2}) may be sharpened:
\begin{thm}\label{WithoutReplWithMalus}
Let $${\cal B}:={N-1\over N-m}\var {Y\over n}={N-n\over N-m}\,{p(1-p)\over n}.$$
If $p_1=\cdots =p_m=p\in(0,1)$, $n<N$, and

(c1) $n_j\leq {3\over 4} N_j$ for $j=1,\ldots,m$, or

(c2) the allocation is proportional, or

(c3) $N_1=N_2=\ldots = N_m=N/m$,

then
$$
\var \sum_{j=1}^m {N_j\over N} {Y_j\over n_j} \geq {\cal B},
$$
i.e.,  the variance of \stratest{}
is at least $(N-1)/( N-m)
\times$ the variance of \simpleest,
and
$$\var \sum_{j=1}^m {N_j\over N} {Y_j\over n_j}
= {\cal B}
\quad\Leftrightarrow\quad
n_j=n/m,\, N_j=N/m\quad\forall j.
$$
\end{thm}
Theorem \ref{WithoutReplWithMalus}(c3) follows from Evans (1951)
(12{\em a, c}).

A special case will illustrate Theorem~\ref{WithoutReplWithMalus}
and  bring out a weak point of the stratification estimator;
the factor $(N-1) /(N-m)$ in~${\cal B}$ appearing here was met 
in (1) (take $n=m$).
Imagine that all strata not only have the same composition,
i.e., $p_j=p$, but also
the same size, i.e., $N_j=N/m$, and that from each stratum
only one  ball is taken, so $n_j=1$ and $n=m$.
Then
$ \sum_{j=1}^m (N_j/N) Y_j/n_j= (1/n)\sum_{j=1}^n Y_j $,
which is distributed as~$X/n$, so that with (\ref{well})
$$
\var \sum_{j=1}^m {N_j\over N} {Y_j\over n_j} = \var {X\over n} =
{N-1\over N-n} \,\var {Y\over n}
={\cal B}.
$$
Splitting the sample over the strata reduces
the
without-replace\-ment bonus from (\ref{well}).

The need for extra conditions in Theorem~\ref{WithoutReplWithMalus}
such as (c1), (c2), or (c3), and the room there is
for Theorem~\ref{WithoutReplWithoutMalus} are demonstrated by considering $m=N_1=n_1=2$,
$n=N-1$, and $N>5$.

Further, one may ask if the condition `$p_1=\cdots =p_m$' is only
the beginning: are there other
distributions of the red balls among
the strata, for which
there are theorems similar to Theorem~\ref{WithoutReplWithMalus}
but with lower bounds that are
even higher than~${\cal B}$ in Theorem~\ref{WithoutReplWithMalus}?
The answer is `no', as long as $N_j:=N/m$ and $n_j:=n/m$
are feasible choices, because for these choices~${\cal B}$ 
is an {\em upper\/} bound (over varying distributions, given $N$, $n$, $m$, and $p$)
for the variance of the stratification estimator
$\sum_{j=1}^m {(N_j/ N)} {Y_j/ n_j}$
 (corresponding 
to
$p_1=\cdots =p_m$; cf. Theorem~\ref{WithoutReplWithMalus}):
\begin{thm}\label{Minimax}
If $N_j=N/m\geq2$, $n_j=n/m$ for all~$j$, and $n<N$
with ${\cal B}$ as in Theorem~\ref{WithoutReplWithMalus}, then
$$
\var \sum_{j=1}^m
{N_j\over N} {Y_j\over n_j} \leq
{\cal B},
$$
i.e.,  the variance of \stratest{}
is at most $(N-1)/( N-m)
\times$ the variance of \simpleest,
and
$$
\var \sum_{j=1}^m
{N_j\over N} {Y_j\over n_j} =
{\cal B}
\quad\Leftrightarrow\quad
p_1=\cdots=p_m.
$$
\end{thm}
Cf. `the worst result to be anticipated' on p.\thinspace 99 of
Evans (1951). (Namely, `for a second' variable of interest; strata that
are good, i.e., different, with respect to the first variable of
interest need not be so for a second.) In the situation
of Theorem~\ref{Minimax}, the worst is not that bad:
in practice, $(N-1)/(N-m)$ will be close to~1, and
it also follows that
{\em if for every stratum the sample size is increased
by~1, the new variance will {not} exceed the simple random
sample variance corresponding to the old sample size.}

%
Theorem~\ref{Minimax} shows that if we want to
curb the badness of stratified sampling and
 proportional allocation is an option,
then it works, at least for
strata of the same size. This makes us realize
that it {\em always\/} works, even
for {\em arbitrary\/} strata sizes, because
independence, (\ref{truth}), and (\ref{PropWithRepl}) imply
$$
\var \sum_{j=1}^m
{N_j\over N} {Y_j\over n_j}
\leq
\var \sum_{j=1}^m
{N_j\over N} {X_j\over n_j}
\leq
{p(1-p)\over n}
$$
under proportional allocation.
Our final results essentially show how the upper bound $p(1-p)/n$ for
$\var \sum_{j=1}^m {(N_j/N)} {Y_j/ n_j} $
can be improved.
\begin{thm}\label{Minimax2}
 If $(N_j/N)n$ and $pN_j$ are integers and $0<pN_j<N_j$ for all~$j$, then
\begin{eqnarray}
\nonumber
{\cal B}&\leq&\min_{\rm Statistician} \max_{\rm Nature} \var
\sum_{j=1}^m {N_j\over N} {Y_j\over n_j}\\
&\leq&{\cal B} + {N-n \over
4(N-m)nN^2}
\sum_{j=1}^{m}{N-mN_j\over N_j-1},
\label{upper bound}
\end{eqnarray}
where `Statistician' means an allocation $(n_1,
\ldots, n_m)$ that satisfies $n_j\leq {3\over 4} N_j$ for all~$j$ or
is proportional,
`Nature' means a distribution $(p_1,\ldots,p_m)$ of the $pN$ red balls
among the
strata (so $p_j\in[0,1]$, $\sum_{j=1}^m p_jN_j = pN$, and $p_j N_j
\in\{0,1,\ldots, N_j\}$), and ${\cal B}$ is as in Theorem~\ref{WithoutReplWithMalus};
in fact, $\max_{\rm Nature}$ does not exceed
 the upper bound in (\ref{upper bound}) if the allocation is proportional.

If $n\leq (3/4)N$, then the upper bound in (\ref{upper bound})
does not exceed $p(1-p)/n$.
\end{thm}

%
%
%
For the lower bound we observe that
it follows from Theorem \ref{WithoutReplWithMalus} and that,
also by Theorem \ref{WithoutReplWithMalus},
if not $N_j=N/m$ for all~$j$,
then `${\cal B}\leq$' may be replaced by `${\cal B}<$'.
%
%
The difference between upper and lower bound  is bounded by $1/4N$
because  $\sum_{j=1}^{m}{(N-mN_j)/( N_j-1)}\leq
\sum_{j=1}^{m}{N/( 2-1)}= mN$; it
reduces to~$0$ in case all strata have
the same size (cf. Theorem~\ref{Minimax}).

The circumstances under which stratified sampling will hurt, have
been called {\em `very unusual'} and {\em `extreme'} (Evans, 1951), {\em `an academic
curiosity'}, which will happen only {\em `mathematically'}
(Cochran, 1963), as well as {\em `quite conceivable'}
(Govindarajulu, 1999).
%
%

\section{Justifications}
\begin{pf*}{Proof of Theorem \ref{WithNotProp}}
By Jensen's inequality we obtain
\begin{eqnarray*}
\var \sum_{j=1}^m {N_j\over N} {X_j\over n_j} &=&
p(1-p) \sum_{j=1}^m {1\over n_j N/N_j} {N_j\over N} \\
&\geq &
p(1-p) {1\over \sum_{j=1}^m {n_jN\over N_j} {N_j\over N  }} \\
&=& {p(1-p)\over n} =\var {X\over n},
\end{eqnarray*}
with `$=$' instead of `$\geq$' if and only if $n_j/N_j$ is
constant, i.e., the allocation is proportional.
\end{pf*}
\begin{pf*}{Proof of Theorem \ref{WithoutReplWithoutMalus}}
Let $m\geq2$, $1\leq n_j\leq N_j$, $N_j\geq 2$, $j=1,\ldots,m$ be integers
with $N=\sum_{j=1}^m N_j$, $n=\sum_{j=1}^m n_j<N$.
In order to prove
\begin{equation}
\label{(1)}
\sum_{j=1}^m {N_j^2\over N_j -1}{N_j - n_j \over n_j} > {N^2\over N-1}{N-n\over n},
\end{equation}
it suffices to prove it for $m=2$.
Indeed, by splitting off one stratum from the urn at a time,
applying (\ref{(1)}) with $m=2$ each time, and observing that
one still has `$\geq$' instead of `$>$' if $n=N$,
 one obtains (\ref{(1)})
for the general case. For  $1\leq k\leq K$ and $1\leq \ell \leq
L$, $K,L>1$, $k+\ell<K+L $ we will prove
\begin{eqnarray}
\label{(2)} {(K+L)^2\over K+L-1}\left( {K+L \over k+\ell}-1\right)
- { K   ^2\over K  -1}\left( {K   \over k  }-1\right)  - {   L
^2\over   L-1}\left( {  L \over \ell}-1\right) < 0.
\end{eqnarray}
To this end we rewrite the LHS of (\ref{(2)}) as $S+T$ with
\begin{eqnarray*}
S={ K+L \over K+L-1}\left( {(K+L)^2 \over k+\ell} -{ K^2 \over k}
- {L ^2\over \ell}\right) = \left( {K+L \over K+L-1}\right) U,
\end{eqnarray*}
\begin{eqnarray*}
T = -{(K+L)^2\over K+L-1}&+&\left( {1 \over K+L-1}-{1\over K-1}\right){K^2\over k}+{K^2\over K-1}\\
 & + & \left( {1 \over K+L-1}-{1\over L-1}\right){L^2\over \ell}+{L^2\over L-1}.\\
\end{eqnarray*}
Note that $\partial U / \partial k = - (K+L)^2 /(k+\ell)^2 + K^2 /
k^2 \geq 0 $ iff $k\leq (K/L)\ell$. Consequently, U is maximal for
$k = (K/L)\ell$ and
\begin{eqnarray*}
U\leq{ 1 \over \ell}\left( {(K+L)^2 \over K / L +1} - KL - L ^2
\right) = 0.
\end{eqnarray*}
Clearly, $T$ is  strictly increasing in~$k$ and~$\ell$ and hence
\begin{eqnarray*}
T < {1 \over K+L-1} \left( -(K+L)^2 + K+L\right)+{K^2 - K\over
K-1}+{L^2-L\over L  -1}=0,
\end{eqnarray*}
which completes the proof.
\end{pf*}
\begin{pf*}{Proof of Theorem \ref{WithoutReplWithMalus}}
The statements corresponding to (c2) and (c3) follow
straightforwardly from the fact that if terms $t_j>0$ have sum
$\sum_{j=1}^m t_j=t$, then for all $a_j\geq 0$, not every $a_j=0$,
we have
$$
\sum_{j=1}^m { a_j\over t_j}=\sum_{j=1}^m \left(\sqrt{a_j\over t_j}\right)^2
                             \sum_{j=1}^m \left(\sqrt{t_j\over  t }\right)^2
\geq {1\over t}\left(\sum_{j=1}^m \sqrt{a_j}\right)^2,
$$
with `$=$' instead of `$\geq$' if and only if
$t_j=t\sqrt{a_j}/\sum_{i=1}^m\sqrt{a_i}$, by Cauchy-Schwarz. With
$a_j=N_j^2$ and $t_j=N_j-1$ we obtain
\begin{eqnarray*}
\sum_{j=1}^m {N_j^2\over N^2} {p(1-p)\over {N_j\over N}n}
{N_j-{N_j\over N}n\over N_j-1}
      &=&{p(1-p)(N-n)\over nN^2} \sum_{j=1}^m {N_j^2\over N_j-1} \\
               &\geq& {p(1-p)(N-n)\over n  N^2  } {N^2\over N  -m},
\end{eqnarray*}
which proves the statements corresponding to (c2),
and with $a_j=1$ and $t_j=n_j$ we obtain
\begin{eqnarray*}
\sum_{j=1}^m{1\over m^2}{p(1-p)\over n_j} {{N\over
m}-n_j\over{N\over m}-1}& = & {p(1-p)\over m^2(N-m)}\left(N\sum_{j=1}^m{1\over n_j}-m^2\right) \\
&\geq&{p(1-p)\over m^2(N-m)}\left(N{m^2\over n} - m^2\right) \\
& = & {p(1-p)\over n}{N-n\over N-m},
\end{eqnarray*}
which proves the statements corresponding to (c3).

In order to prove the statements corresponding to (c1), we observe
that the function
$$
\psi(x,y)={{1\over y}-1\over 1-x}
$$
 is strictly
convex on $(0,1)\times (0,{3\over 4}]$,
because it is strictly convex on any segment in
$(0,1)\times (0,{3\over 4}]$. On any segment $\{t(x_1,{3\over 4}) + (1-t) (x_2, {3\over 4})\}$, namely, the strict
convexity is clear, while for a segment not contained in $y={3\over 4}$
we have
$$
\left\vert\matrix{
{\partial^2\over \partial x^2} \psi (x,y) &
   {\partial^2\over \partial x\partial y} \psi (x,y)\cr
{\partial^2\over \partial y\partial x} \psi (x,y) &
   {\partial^2\over \partial y^2} \psi (x,y)\cr
}\right\vert
=
\left\vert\matrix{
{2(1-y)\over y (1-x)^3}  &  {-1\over y^2 (1-x)^2} \cr
{-1\over y^2 (1-x)^2}  &  {2\over y^3(1-x)} \cr
}\right\vert
= {3-4y \over y^4(1-x)^4}.
$$
The Hessian of~$\psi$, therefore, of which the determinant is the
product of the eigenvalues and the sum of the diagonal elements
is the sum of the eigenvalues, is positive
definite outside $y={3\over 4}$, so the second derivative of
$t\in[0,1]\mapsto \psi \bigl( t(x_1,y_1) + (1-t)(x_2,y_2)\bigr)$
is positive on $(0,1)$.

Consequently, applying Jensen's inequality to the random
2-vector $\pmatrix{X\cr Y}\colon j\in\{1,\ldots,m\}\mapsto
\pmatrix{1/N_j\cr n_j/N_j}\in\R^2$ with
$P(\{j\})={N_j/ N}$, $j=1,\ldots,m$,
gives
\begin{eqnarray*}
\sum_{j=1}^m {{1\over n_j/N_j}-1\over 1-1/N_j}\, {N_j\over N}
& = & E\psi (X,Y)\geq \psi (EX,EY)\\
& = & \psi ( {m\over N},{n\over N}) = {N\over N-m}\left({N\over
n}-1\right),
\end{eqnarray*}
with `$=$' instead of `$\geq$' if and only if $1/N_j$ and $n_j/N_j$
are constant. This proves the statements corresponding to condition (c1).
\end{pf*}
\begin{pf*}{Proof of Theorem \ref{Minimax}}
Under $n_j=n/m$, $N_j=N/m$ the variance of the stratification
estimator becomes
$$
\sum_{j=1}^m{1\over m^2}{p_j(1-p_j)\over {n\over m} }
             {{N\over m}-{n\over m}\over{N\over m}-1}
$$
and $\sum p_jN_j = pN$ becomes $\sum p_j=mp$, while by Cauchy-Schwarz
$$
\sum p_j(1-p_j)= mp - {\sum p_j^2\cdot \sum 1^2\over m}
\leq
mp -{1\over m}\left(\sum(p_j\cdot 1)\right)^2 = mp(1-p).
$$
This proves Theorem \ref{Minimax}. If,
in the situation of Theorem \ref{Minimax},
for every stratum the sample size is increased by~1,
we have
$n_{\rm new}=n+m$ and the new variance 
will not exceed
$$
{\cal B}_{\rm new}
={N-m-n\over N-m}\,{p(1-p)\over n+m}\leq{N-n\over N-1}\,{p(1-p)\over n}.
$$
\end{pf*}
\begin{pf*}{Rest of Proof of Theorem \ref{Minimax2}}
For the upper bound, let $\alpha_j, \beta_j, 1 \leq j \leq m,$ be positive reals with
$\sum_{j=1}^m \beta_j =1.$ By the multiplier method of Lagrange we
see that $\sum_{j=1}^m \alpha_jp_j(1-p_j)$ attains its maximum
over $p_j 
$
under the side condition $\sum_{j=1}^m\beta_j(p_j - p)=0$ at
$$
p_j= {1\over 2} + {\beta_j(p-{1\over 2})
\over \alpha_j \sum_{k=1}^m \alpha_k^{-1}\beta_k^2 } 
$$
with maximum value equal to
$$
{1\over4}\left(\sum_{j=1}^m \alpha_j - {(2p-1)^2 \over
\sum_{k=1}^m \alpha_k^{-1}\beta_k^2 }\right).
$$
With $$\alpha_j= {N_j^2(N_j-n_j) \over N^2 (N_j -1)n_j}, \quad
\beta_j={N_j\over N},\quad 1\leq j \leq m,
$$
this shows that under proportional allocation, $n_j=(N_j/N)n$,
\begin{equation}\label{a}
\max_{\rm Nature} \var \sum_{j=1}^m {N_j\over N} {Y_j\over n_j}
\leq {N-n \over 4(N-m)n} \left({N-m \over N^2}\sum_{j=1}^m {N_j^2
\over N_j -1} - (2p-1)^2\right)
\end{equation}
holds. The right-hand side of (\ref{a}) is equal to
the upper bound in (\ref{upper bound}).

Finally, as
$${\cal B}={N-n\over N-m}\,{p(1-p)\over n},$$ the fact that the upper bound in
(\ref{upper bound}) does not exceed $p(1-p)/n$ is equivalent to
$$
{N-n\over 4(N-m)nN^2}\sum_{j=1}^m {N-mN_j\over N_j -1}
\leq
{n-m\over N-m}{p(1-p)\over n}.
$$
Suppose $N_1\leq N_2\leq\cdots\leq N_m$. As $1\leq pN_1\leq {N_1-1}$,
we have
$$
{n-m\over N-m}{p(1-p)\over n}\geq
{n-m\over (N-m)n}{{1\over N_1}\left(1-{1\over N_1}\right)}.
$$
Consequently, it is sufficient to prove
$$
\sum_{j=1}^m {N-mN_j\over N_j -1}
\leq
{4(N-m)nN^2\over N-n}\cdot{(n-m)(N_1-1)\over (N-m)n N_1^2},
$$
whose left-hand side is equal to the left-hand side in
$$
(N-m)\sum_{j=1}^m { 1 \over N_j -1}-m^2
\leq
(N-m) {m\over N_1 -1}-m^2 = {m(N-mN_1)\over N_1 - 1}
$$
(remember $N_j\geq N_1$), so that it is sufficient to prove
$$
m(N-mN_1)\left( {N_1\over N_1 - 1}\right)^2
\leq
4(n-m) {N^2\over N-n}.
$$
This is true if
$$
m\left(N-m{N\over n}\right)\left( {N/n\over N/n - 1}\right)^2
\leq
4(n-m) {N^2\over N-n}
$$
(as $N_1\geq N/n$), which is equivalent to
$$
mN(n-m)
\leq
4n(N-n)(n-m) ,$$
which is true if $n\leq (3/4)N$, because $m\leq n$.
\end{pf*}
\Blind
{\bf Acknowledgements}

Andries Lenstra was supported by The Netherlands Foundation for
Scientific Research (NWO), by EURANDOM, Eindhoven, the
Netherlands, by Texas Tech University, Lubbock, TX, U.S.A., and by
Oberlin College, Oberlin, OH, U.S.A. \EndBlind

\vfil
\author{Chris A.J. \surname{Klaassen}}

\institute{Korteweg-de Vries Institute for Mathematics,
Universiteit van Amsterdam,
Plantage Muidergracht 24, 1018 TV Amsterdam, the Netherlands,
 chrisk@science.uva.nl}
\author{Andries J. \surname{Lenstra}}

\institute{Korteweg-de Vries Institute for Mathematics,
Universiteit van Amsterdam,
Plantage Muidergracht 24, 1018 TV Amsterdam, the Netherlands,
 andriesl@science.uva.nl}

\eject
\end{document}